# STATUS REPORT AND FUTURE PLANS FOR THE PEP-II *B* FACTORY*

Alan S. Fisher, SLAC, Stanford, CA 94309, USA

*Abstract*

The PEP-II B Factory at SLAC has been in operation with the *BABAR* detector since the summer of 1999. The peak currents and luminosity steadily increased through the end of the 2000 run on October 31. By that time, 0.75 A of electrons (the design current) routinely collided with 1.4 A of positrons in 657 bunches, to give a peak luminosity of $2.6 \times 10^{33}$ cm$^{-2} \cdot$s$^{-1}$. We delivered an integrated luminosity of up to 184 pb$^{-1}$ per day and 1033 per week; *BABAR* logged a total of 25 fb$^{-1}$. Three days of machine development at the end of the run raised the peak luminosity to $3.29 \times 10^{33}$, exceeding the design goal of 3.0. *BABAR* then logged data for an hour, starting with a peak of 3.20. Next, we achieved the design value for positron current, 2.14 A, operating without collisions.

The major limitation has been growth in the size of the positron beam in the low-energy ring (LER) due to electron clouds and multipacting. Since the arcs benefit from antechambers and a TiN coating with low secondary emission, our efforts have concentrated on the straights, where we added solenoid windings. Each straight wound allowed higher LER current without blow-up, and a consequent increase in luminosity. Measurements of the bunch-by-bunch luminosity showed that the effect becomes significant by about the tenth bunch in a train, but clears after a short gap. Careful control of the fill pattern has thus been essential in raising luminosity.

For the 2001 run, which began in February, we added a third LER RF station, to collide with more LER current and more bunches, and also tried reducing $\beta_x^*$ from 50 to 35 cm, with the goal of reaching a luminosity of $5 \times 10^{33}$ by year's end. In 2003, we plan to add a fourth LER and sixth HER RF station to reach $1 \times 10^{34}$.

**Table 1.** Some PEP-II Design Parameters.

|  | LER | HER |
|---|---|---|
| Circumference [m] | 2199.322 | |
| RF frequency [MHz] | 476.00 | |
| Harmonic number | 3492 | |
| Colliding bunches | 1658 | |
| Current [mA] | 2140 | 750 |
| Beam energy [GeV] | 3.119 | 8.973 |
| CM energy [GeV] | 10.58 | |
| $\beta_x^*$, $\beta_y^*$ [cm] | 50, 1.5 | 50, 1.5 |
| $\varepsilon_x^*$, $\varepsilon_y^*$ [nm] | 49, 1.5 | 49, 1.5 |
| $\sigma_x$, $\sigma_y$ [µm] | 157, 4.7 | 157, 4.7 |
| $\sigma_z$ [mm] | 12.3 | 11.5 |
| Tune shift | 0.03 | |
| Aspect ratio at IP (*v/h*) | 0.03 | |
| Crossing angle | 0 (head on) | |

## 1 INTRODUCTION

The PEP-II *B* Factory, a 2.2-km asymmetric collider at the Stanford Linear Accelerator Center[1], was built in collaboration with the Lawrence Berkeley[2] and Lawrence Livermore[3] National Laboratories to study *CP* violation by tracking decays of *B* mesons moving in the lab frame. At a single interaction point (IP), 9-GeV electrons in the high-energy ring (HER) collide at zero crossing angle with 3.1-GeV positrons in the low-energy ring (LER). Table 1 lists some PEP design parameters[4]. The first collisions were observed in July 1998, when the rings were commissioned without the *BABAR* detector. After its installation in May 1999, commissioning of the full system began; physics runs followed shortly afterward. By the end of 1999, the luminosity had passed $1 \times 10^{33}$ cm$^{-2} \cdot$s$^{-1}$.

## 2 THE 2000 RUN

### 2.1 Highlights

The 2000 run, from January through October, achieved a number of records, as Table 2 shows. In the machine-development days at the end of the run, the luminosity reached a peak of $3.29 \times 10^{33}$ cm$^{-2} \cdot$s$^{-1}$, exceeding the design goal of 3.0. *BABAR* then took data for a short while at 3.20. This luminosity was achieved with currents of 1550 mA in LER and 800 mA in HER, filling only 692 bunches and so suggesting that higher luminosities will be possible in the future. Table 3 summarizes the parameters for this luminosity.

**Table 2.** PEP-II Records, as of March 2001. The data incorporates the new luminosity calibration discussed in Section 2.1.

|  | Goal | Achieved |
|---|---|---|
| Peak LER current [mA] | 2140 | 2140 |
| Peak HER current [mA] | 750 | 920 |
| Number of full buckets | 1658 | 1658 |
| Peak luminosity [$10^{33}$ cm$^{-2} \cdot$s$^{-1}$] | 3.0 | 3.29 |
| Peak luminosity with *BABAR* [$10^{33}$ cm$^{-2} \cdot$s$^{-1}$] | 3.0 | 3.20 |
| Integrated luminosity per shift [pb$^{-1}$] | 45 | 66 |
| Integrated luminosity per day [pb$^{-1}$] | 135 | 184 |
| Integrated luminosity delivered per week [pb$^{-1}$] | 785 | 1033 |
| Integrated luminosity delivered per month [pb$^{-1}$] | 3300 | 3820 |
| Total luminosity integrated by *BABAR* [fb$^{-1}$] |  | 26 |

* Work supported by the US Department of Energy under Contract No. DE-AC03-76SF00515.

**Table 3.** Parameters for Record Luminosity.

| | LER | HER |
|---|---|---|
| Peak luminosity [$10^{33}$ cm$^{-2}$·s$^{-1}$] | 3.29 | |
| Current [mA] | 1550 | 800 |
| Number of bunches | 692 | |
| $\Sigma_{x,y}$ at low current [µm] | 210, 6.7 | |
| IP beam sizes, *x, y* (average) [µm] | 147, 5 | |
| Horizontal tune shift | 0.069 | 0.060 |
| Vertical tune shift | 0.055 | 0.028 |

During October, the integrated luminosity, measured over a shift, day, week or month, all exceeded the design goals. Also, in single ring operation, the LER current was raised to the design value of 2140 mA, and the HER reached 920 mA, exceeding the goal of 750.

This luminosity is a bit higher than values reported previously. The BABAR group recently refined their calibration of our luminosity monitor, based on a detailed off-line study of $e^+e^- \to e^+e^-$ and $e^+e^- \to \mu^+\mu^-$ decays. This correction, which adds 6% to the luminosity, is incorporated here.

## 2.2 Thermal and Vacuum Issues

As the currents increased during the run, thermal limitations became apparent in several components. A number of bellows required modest air cooling. More severely, the connectors and directional couplers on the longitudinal feedback kickers overheated, at one point burning some of the Heliax-cable insulation. New couplers and additional cooling were added immediately, with more improvements added during the downtime (Sec. 3.4); temperatures there are now closely monitored with thermocouples at numerous points.

Two permanent-magnet dipoles (B1), located inside BABAR on either side of the IP, bring the beams into head-on collision. They create hot synchrotron fans that travel outside the detector to cooled chamber surfaces 10 to 15 m away. The hottest fan, from the electrons, strikes the long "high-power dump" chamber at grazing incidence over several meters. This water-cooled, rectangular, copper chamber began to leak at one corner, where it was brazed to a stainless-steel vacuum flange. The hot fan caused one side of the chamber to expand more than the opposite side; over many cycles, the stress cracked the braze. While building a replacement chamber with an improved design, we patched the leak to finish the run. A metal "boot" was epoxied to the region outside the leak and evacuated with a turbomolecular pump, in order to maintain acceptable vacuum inside the beampipe. However, we limited the electron current to a maximum of 600 mA until the final weeks of the run, in order to avoid further damage. The chamber was replaced during the downtime between the 2000 and 2001 runs.

For the longer term, the most worrisome thermal constraint has been found on the bellows at the forward (downstream electron) end of the "support tube"—the 40-cm-long beryllium beampipe enclosing the IP and running between the inboard ends of the two B1 magnets. Some power from the beam goes into a mode trapped by a synchrotron-radiation mask just outboard of the support tube, and this power appears to leak through the shielding fingers of the bellows. The water lines for cooling the support tube enter from the other end and turn around a few centimeters away from the bellows, reducing the cooling there. The bellows is deep inside the detector, tightly surrounded by the silicon vertex tracker (SVT), making it inaccessible until the scheduled access to the SVT in two years. The most we can do until then is to force a small flow of dry air along the support tube. The benefit is limited, since a radial ion pump outboard of the bellows on each side blocks the flow of air along the tube wall.

Measurements of the dependence of the bellows temperature on the currents in the two rings show a largely quadratic behavior, indicative of heating from high-order modes rather than synchrotron radiation. The LER also has a linear term. We interpret this as a change in bunch length with current, since we also found a linear dependence on bunch length by varying the LER's RF voltage. The cross term from the product of the HER and LER currents, responsible for about 10% of the heating, varied with a scan of the arrival phase of the HER relative to the LER at the IP. The spatial periodicity suggested a trapped mode with a frequency of 5.4 GHz.

## 2.3 Multipacting and Electron Clouds

Last spring and summer, the current in many LER ion pumps grew slowly as a function of beam current, then crossed a threshold and increased at a much greater rate. We also observed that the specific (normalized) luminosity, which depends only on the sizes of the overlapped beams at the IP, was found to drop significantly at currents below those at which beam-beam limitations should come into effect.

Although pump current is normally proportional to pressure, much of this increased current dropped immediately when the beam was aborted. The rest of the signal, from gas desorbed by the beam, decayed exponentially over tens of seconds. We attributed the rapidly lost pump current to the collection of electrons released by synchrotron radiation from the beampipe walls. The rapid rise with current then was interpreted as a multipactor process: the electrons are drawn into the positron beam and accelerated across the beampipe, releasing more electrons in a cascade that creates an electron cloud. In some cases the process was clearly resonant: as the beam current increased over threshold, the pump current first increased rapidly to a peak, then decreased. The details of this resonance depend on the bunch current, bunch spacing, beampipe geometry, and wall cleanliness. A solenoid wrapped around the beampipe adjacent to the ion pump reduced the electron cloud by preventing photoelectrons from crossing through the beam potential and ejecting more electrons from the opposite wall.

At the same time, we observed that the LER beam size, measured by the synchrotron-light monitor, exhibited similar behavior, growing little with current below a threshold, and then increasing rapidly. The threshold was lower, and the growth somewhat more rapid, with the beams in collision, but it was also present in single-beam operation. We attribute it to the onset of the electron-cloud instability (ECI)[5] due to the cloud created by multipacting. Two techniques have been helpful in controlling this effect: solenoidal fields and short gaps in the fill pattern.

Since the arcs have much more synchrotron radiation than the straights, they might be expected also to have denser electron clouds. However, LER arc chambers are made of aluminum extrusions with an antechamber. Downstream of each dipole, a "photon stop" in the antechamber receives the hard synchrotron radiation from the bend. Secondary electrons are thus kept out of the main chamber and the field of the positron beam. To further suppress the ECI, the aluminum walls are coated with titanium nitride (TiN), to lower their secondary-emission coefficient. In contrast, the beampipes in PEP's straight sections are mostly stainless-steel cylinders. The material and the circular geometry permit the growth of the instability.

Consequently, we added our solenoids to the straights. Solenoidal windings wrapped directly on the tubes, with a typical field of 30 G, noticeably helped in reducing the LER beam blow-up, even when only a single straight was wrapped with about 100 m of windings in the drift spaces. We have now added solenoids to all six straights, for a total of approximately 600 m. As each straight was wound, the threshold for beam-size growth moved higher and the luminosity increased.

The electron cloud builds up quickly, but also decays quickly, and so small gaps in the fill pattern can have a big effect. PEP's design called for filling every second RF bucket (2.1 ns per bucket) around the entire ring except for a 5% gap for both clearing ions and ramping up the beam-abort kicker's field. However, our injection control allows us to fill any specified bucket to any desired charge. As we gradually raised the beam currents while optimizing luminosity, we began with fewer buckets, to keep a high charge per bunch until limited by beam-beam effects. However, we kept to uniform fill patterns; early in the year we filled every $8^{th}$ (called a "by-8" fill), then every $6^{th}$, $4^{th}$, and finally $3^{rd}$ by October. (We preferred not to use by-2 too quickly because bunch $n$ then experiences a parasitic near-crossing with bunch $n\pm1$ on either side of the IP, just outboard of the B1 magnets.)

Our luminosity monitor has a mode that sweeps a gate across the fill pattern to provide a display of the luminosity from each bunch. We observed a strong drop (≈60%) in luminosity from the first to the last buckets, and this drop had a time constant of about 100 buckets. By inserting various "microgaps" in the fill pattern, we determined that removing even 6 to 12 buckets allows a noticeable recovery. The luminosity significantly improved after adjusting the total number of filled buckets, the number of and length of gaps, and the length of the individual trains. The record luminosity was achieved with a by-3 fill in which 10 bunches were filled, followed by 6 left empty (bucket numbers 0, 3, 6…27; 48, 51…), for a total of 692 bunches.

The bunch-by-bunch luminosity and current monitors also provide insight into the relative strengths of the two beams. The first LER train, following the main gap, is better focused than subsequent ones because it hasn't encountered a thick electron cloud. As a result, there can be significant loss of electrons from the first HER train. After top-off, this train begins with the highest luminosity, which drops as electrons are lost. To counteract this, we gradually increase the charge per LER bunch in the first train with a programmed ramp.

## 2.4 Other Issues

Considerable effort went into learning how to decouple the rings and compensate for the twist given to the beams by *BABAR*'s solenoidal field. The most helpful technique was to generate a large orbit wave with a corrector in one plane, then to observe the coupling into the other plane with the beam-position monitors (BPMs). Two correctors 90º apart in betatron phase are needed per plane. Careful comparisons with the model can localize errors in the compensation provided by skew quadrupoles in the region of the IP.

A number of feedbacks were added during the run to keep the luminosity high. The IP loops dither the HER's position ($x$, $y$) and angle ($x'$, $y'$) at the IP to maximize the luminosity monitor's signal. HER/LER combined angle loops move both rings together at the IP to optimize pointing at the luminosity monitor, about 10 m away. Other loops maintain beam position in sextupoles and in the pick-ups for transverse feedback.

HER and LER orbit loops use BPMs in the arcs on either side of the IP to determine a kick near the IP that compensates for thermal motion of the magnet supports ("rafts") next to BABAR, mostly due to diurnal temperature variations. Since that time, we have had some success in reducing the motion by using our temperature-controlled cooling water to limit temperature changes in the rafts on either side of BABAR, and we are trying to determine if these loops remain helpful.

Backgrounds in BABAR remain a problem. Radiation-detecting PIN diodes inside BABAR, near the SVT, abort the beams several times a day when the levels exceed a safe dose. We have been learning how to tune the rings to reduce backgrounds. Improvement should be seen later in the 2001 run (once sufficient scrubbing has taken place) due to our downtime work on the IP-area vacuum chambers.

The HER is also subject at times to sudden drops in lifetime, often accompanied by bursts of radiation at the SVT diodes, leading to a beam abort. We tentatively attribute this to dust particles falling into the beam, but have not been able to trace the source of the dust to individual vacuum chambers. At times the lifetime remains low (and the backgrounds high) for several minutes, which may be caused by a plasma from ionized dust. At times shaking the HER (with swept-frequency sinusoidal motion excited through the transverse-feedback kickers) restores the lifetime, but more often we are forced to abort the beam.

## 3 2000–2001 DOWNTIME WORK

The PEP shutdown—from November 1 through February 2—allowed us to make several improvements addressing the issues discussed above.

### 3.1 Solenoids for the Arcs

We have begun to wrap the far longer extent of the arcs, to determine whether the arcs contribute in part to the ECI. Half of one arc was wound in January. The work is more difficult because the antechamber gives the beampipe a greater perimeter, and because a steel support tube running above the aluminum chamber makes wrapping difficult.

The beam chamber within the extrusion is an ellipse, with a narrow opening on the +$x$ side leading through a neck to the antechamber. With this geometry, it may not be necessary to use a solenoidal field; instead, a horizontal dipole field oriented to direct stray electrons through the neck could be sufficient to suppress the instability. Such a field turns out to be much easier to arrange on our chambers. Long L-shaped forms can be wound in advance with coils running along the length of the L. The base of the L, with one side of the coil, is placed in contact with the chamber, while the returning side of the coil runs along the other arm of the L, further from the beam. When placed above and below the elliptical part of the beampipe, the pair of coils produces a horizontal field at the beam. The polarity of the field is reversed every 2 m to cancel the steering. We are installing half an arc of these coils for comparison with half an arc of solenoids.

### 3.2 Vacuum Improvements

The HER's high-power dump chamber (Sec. 2.2) was replaced, eliminating the major vacuum leak near the IP. Starting 12 m away from the IP on the opposite (outgoing electron) side, a new chamber with large titanium-sublimation pumps was installed to lower the pressure in a 9-m region that appears to be a significant source of backgrounds.

The number of ion pumps was doubled in LER Arc 7, following a program to increase the base pumping speed in all the LER arcs, and so to obtain a higher beam lifetime, which has typically been 1 hour when colliding. (Ti-sublimation pumps in the antechambers provide additional speed for the outgassing from the photon stops.) The leaks that caused a somewhat elevated pressure were fixed. A few minor leaks remain due to difficulties with the tin-foil seals on the aluminum flanges.

### 3.3 Background Reduction

New horizontal collimators were installed in the LER, 12 and 25 m upstream of the IP. These have a motorized jaw on both sides ($\pm x$).

A shielding wall was installed at the tunnel mouth, on the forward side of BABAR, to reduce backgrounds due to radiation from the tunnel. A similar wall was previously installed on the backward side.

We installed pulsed beam-separator magnets, which should provide cleaner injection by separating the beams during fills and top-offs, and then rapidly bringing them back into collision.

### 3.4 RF and Feedback

A third LER RF station was added, to provide enough power for currents of up to 3 A. The cooling for the connectors and couplers of the longitudinal feedback kickers was improved, as discussed in Sec. 2.2.

*3.5 Survey*

The entire ring was resurveyed. Errors in individual components were found and corrected. In addition, we adjusted for some settlement of BABAR by smoothly adjusting the beamlines on either side to match, since there was no time to reposition the detector.

## 4 STATUS OF THE 2001 RUN

The goal for the run, which lasts from February to the end of August, is to reach a peak luminosity of $5\times10^{33}$ cm$^{-2}\cdot$s$^{-1}$, and to deliver an integrated luminosity of 32 pb$^{-1}$ over this time.

At this time (mid-March), the run has been difficult, probably because of many changes in too short a time. In January, even before the run started, the RF frequency was increased by 600 Hz, to better center the beams in the rings. While this should be beneficial in the longer term, we then weren't able to return to the horizontal orbits of October as we turned back on.

While the machine was still scrubbing, recovering from the extensive vacuum work, we took the opportunity to do machine development, and lowered $\beta_x^*$ from 50 to 35 cm, first in the LER, and then the following week in the HER.

With the tighter beam in LER, the positrons for the first time dominated the electrons in beam-beam. Electron beam loss was common, and there was little margin for error in tune adjustment with changes in currents. However, we were able to get to a peak luminosity of $2\times10^{33}$. The subsequent decrease in the HER $\beta_x^*$ strengthened HER somewhat, but the tune space remained difficult, and beam loss was unacceptably frequent, precluding effective tuning.

Early in March we changed course, trying to revert as much as possible to the configuration of late October. We did not, of course, undo the magnet repositioning from the new survey, but we restored the RF frequency and the $\beta_x^*$ values. The magnet configurations were reloaded, and the ring orbits were steered back to the saved orbits from October.

After these changes, it was easier to maintain the beams in collision, providing time to tune the IP skew quadrupoles and optimize luminosity. The peak luminosity is typically $2.2\times10^{33}$ at this time.

Thermal constraints now limit the currents and hence the luminosity. When new hot components are discovered, we add more thermocouples and set the beam-abort thresholds with caution. The result is that thermal aborts are now more frequent than before, not because the machine is hotter, but because we are looking in hotter places and have set lower limits. For example, the connectors on the kicker chambers for longitudinal feedback still get hot and now have numerous thermocouples. The heating is more severe with the by-3 fill pattern used in October (when we were less aware of the problem), leading us to operate mostly with by-2 now.

In addition to the hot components found last year, we have recently discovered a high temperature that comes and goes erratically at a LER gate valve 12 m before the IP. The cause may be a LER synchrotron fan from upstream that should be blocked by a mask. However, the mask does not span the full height of the chamber, and small changes in the orbit may allow some of the fan to hit the valve.

If the vertical angle of the HER at the IP is high, then its B1 fan heats a large HER vacuum flange about 8 m away. Because this fan caused a leak in the past, the temperature limit is now set low. The HER IP position and angle feedback loops move HER to find the best overlap with LER for luminosity. At times the feedbacks appear to move the HER angle upward, apparently following some change in the LER and causing an abort. We suspect an intermittent problem with a BPM may cause a LER orbit feedback to move, potentially causing either of these thermal problems. Consequently, we are reviewing the performance of all loops. Some may need improvement, while others may no longer be needed, since the drifts they were designed to compensate may now be reduced.

We are also losing beam several times a day due to trips of the HER RF stations. In some events, the RF may not have been the cause; instead, the high reflected power appears to follow a beam abort from some other source that fails to register with the abort system.

Once we have fully returned to October's performance, we plan to reintroduce the changes in frequency and $\beta_x^*$. When we commission the third LER RF station, the shorter bunch length will permit us to reduce $\beta_y^*$ from its present 12.5 mm; a 10-mm lattice is ready to try with beam, and a 7-mm lattice is ready for a magnet test. We will also commission the pulsed separator magnets.

## 5 PLANS FOR 2002 AND BEYOND

The medium-term goal is to reach a luminosity of $10^{34}$ cm$^{-2}\cdot$s$^{-1}$ by the end of 2003. We have begun building two more RF stations (one per ring), to get the higher current and shorter bunches this goal requires. The new stations should be ready by the fall of 2002.

For the long term, we have begun studying paths to $3\times10^{34}$ cm$^{-2}\cdot$s$^{-1}$. Both rings would require more current—1.5 A in the HER and 4 A in the LER—and thus more cooling and more RF power. Similarly, the bunch-by-bunch feedbacks would have to be strengthened. We would further reduce $\beta_y^*$ to about 6 mm, which would involve moving the IP quadrupoles inward. We would then need to further shorten the bunches, perhaps with a lower $\alpha$ or with third-harmonic RF cavities. The new, tighter IP would change from our head-on configuration to one with a crossing angle of perhaps ±3 mrad, allowing more bunches without the parasitic crossings.

## 5 ACKNOWLEDGMENT

I wish to thank the accelerator and *BABAR* physicists, machine operators, and members of SLAC's technical staff, whose efforts have made PEP-II successful.